\documentclass[pss]{wiley2sp} % provides pss two-column style
\usepackage{amsmath}
\begin{document}

% Title of the article
\title{Levitons for electron quantum optics}

% Abbreviated title for the page headers
\titlerunning{levitons }

% Authors
\author{%
  %First Author\textsuperscript{\Ast,\textsf{\bfseries 1}},
  D. Christian Glattli\textsuperscript{\Ast,\textsf{\bfseries 1}},
  Preden S. Roulleau\textsuperscript{\textsf{\bfseries 1}},
  }
% Abbreviated list of authors for the page headers
\authorrunning{D.C. Glattli et al.}

%E-mail-address of corresponding author
\mail{e-mail
  \textsf{christian.glattli@cea.fr}
  }
% author's affiliations/addresses
\institute{%
  \textsuperscript{1}\,Nanoelectronics Group, SPEC, CEA, CNRS, Universit\'{e} Paris-Saclay, CEA-Saclay, 91191 Gif-sur-Yvette, France }%\\

\received{XXXX, revised XXXX, accepted XXXX} % do not change, will be filled in by the publisher
\published{XXXX} % do not change, will be filled in by the publisher

% Please select about four verbal keywords for your manuscript.

\abstract{%
% This is a macro for the typesetting of two-column text in an
% abstract. It will typeset the two arguments in \abstcol{}{} as the
% left and right column inside the abstract box. At the
% columnbreak there will be always a columnbreak (\par), so both
% columns start with a new paragraph. No automatic column height
% balancing is done.
%
% If used with a \titlefigure it will silently output both
% parameters as consecutive paragraphs.
%
% The macro is defined exclusively inside the argument of \abstract{};
% if used outside it will raise an error.
%
% Usage: \abstcol{<left column>}{<right column>}
\abstcol{%
  Single electron sources enable electron quantum optics experiments where single electrons emitted in a ballistic electronic  interferometer plays the role of a single photons emitted in an optical medium in Quantum Optics. A qualitative step has been made with the recent generation of single charge levitons obtained by applying Lorentzian voltage pulse on the contact of the quantum conductor. Simple to realize and operate, the source emits electrons in the form of striking minimal excitation states called levitons. We review the striking properties of levitons and their possible applications in quantum physics to electron interferometry and entanglement.

  }{%
  }}

% The class file requires the standard graphicx Latex package. See the 'LaTeX
% standard graphics and color packages documentation' for more information at
% <http://tug.ctan.org/tex-archive/macros/latex/required/graphics/grfguide.pdf>.
%
% Accepted figure file formats depend on which LaTeX flavour is used.
% Classic LaTeX is always able to use Encapsulated Postscript (EPS);
% PDFLaTeX can't use this but accepts PDF, JPG, PNG, and GIF formats.
%
% See examples for implementing graphics in floating figure environments later in this file.
% If \titlefigure is given, it takes as its mandatory parameter the
% name (without extension) of some figure file.
\titlefigure[clip, width=7cm, height=6.1cm]{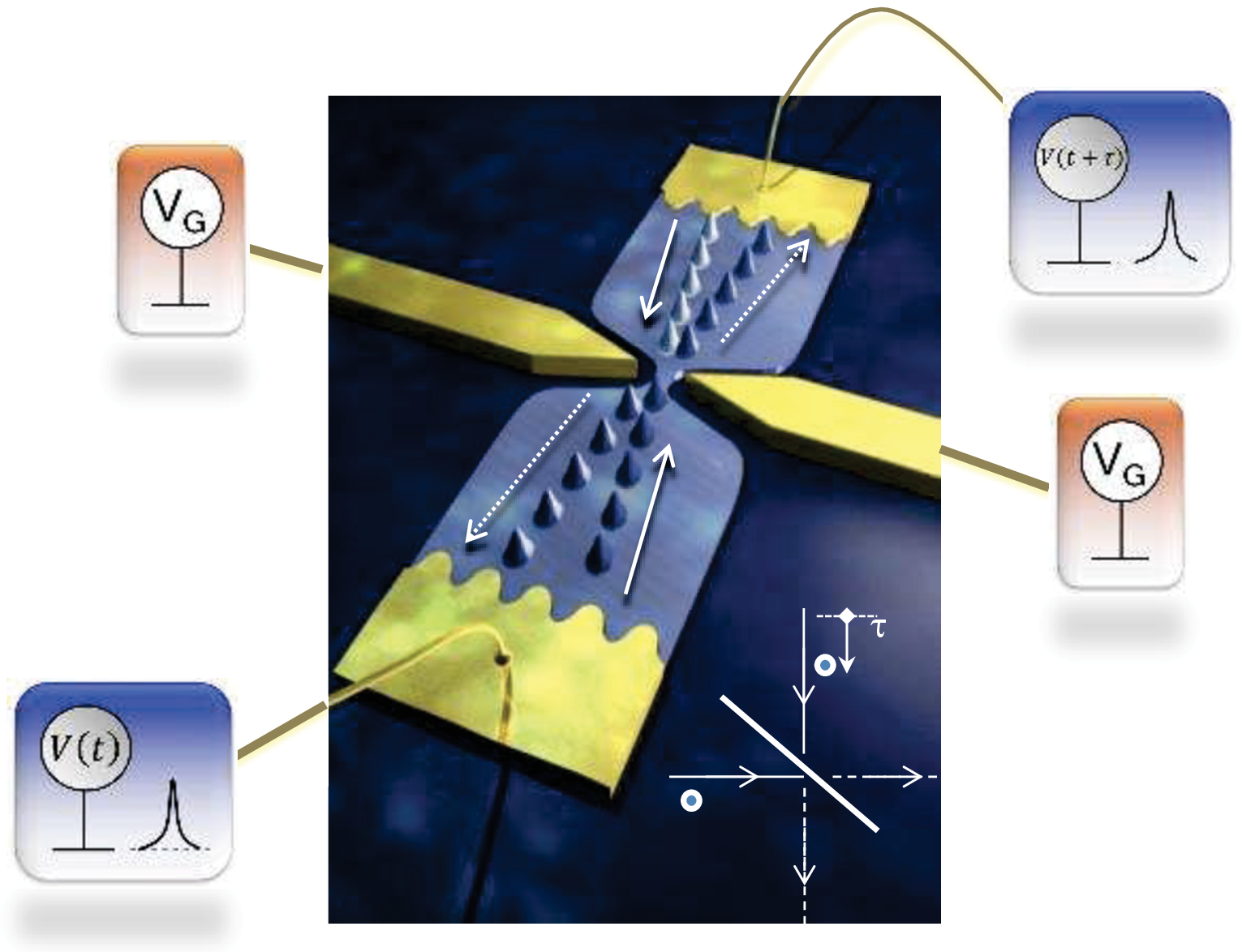}
\titlefigurecaption{%
  Schematic generation of time resolved single charges called levitons using Lorentzian voltage pulses applied on a contact. A Quantum Point Contact is used to partition the levitons for further analysis. Injecting levitons on opposite contacts with a delay $\tau$ enables to probe electronic like Hong Ou Mandel correlations.}

\maketitle   % please do not remove

\section{Single electron sources}
In this introduction, we will distinguish single charge sources from coherent single electrons sources. The former have been developed for quantum metrology where the goal is to transfer an integer charge at high frequency $f$ through a conductor with good accuracy to realize a quantized current source whose current $I=ef$ shows metrological accuracy. The latter, the coherent single electrons source, aims at emitting (injecting) a single electron whose wave-function is well defined and controlled to realize further single electron coherent manipulation via quantum gates. The gates are provided by electronic beam-splitters made with Quantum Point Contacts or provided by electronic Mach-Zehnder and Fabry-Pérot interferometers. Here it is important that the injected single electron is the only excitation created in the conductor. The frequency $f$ of injection is not chosen to have a large current, as current accuracy is not the goal, but only to get sufficient statistics on the electron transfer events to extract physical information.

\subsection{single charge sources for current standards}
The first manipulation of single charges trace back to the early 90's where physicists took advantage of charge quantization of a submicronic metallic island nearly isolated from leads by tunnel barriers. The  finite energy $E_{C}=e^{2}/2C$ to charge the small capacitor $C$ with a single charge being larger than temperature (typically one kelvin for $C=$1fF), Coulomb Blockade of tunneling occurs allowing for single charge manipulations. An architecture made of several island in series lead to single electron pumps and turnstiles \cite{Geer90,Poth91}. Pumps and turnstiles were dedicated to quantum metrology with the hope to realize a current standard where the current $I=ef$ is related to the frequency of single charge transfer through the device. A 7-junction pumps lead soon to $15.10^{-8}$ current accuracy but limited to too low currents \cite{Kell96} for metrology. Later hybrid superconducting normal junctions single electron pumps lead to higher operating frequencies to reach $10^{-6}$ accuracy \cite{Peko13}.

In parallel, electron pumps have been realized using 2D electrons confined in GaAs/GaAlAs semiconductors heterojunctions where the metallic island is replaced by a quantum dot obtained by surface gate depletion. Here pairs of split gates were used to form a narrow constriction which selects the number of electronic modes transmitted between the dot and the leads and ultimately leads to a 1D tunnel barrier. This Quantum Point Contact (QPC) offers tunability of the tunnel barrier whose fast manipulation was exploited to capture a single electron in the dot from the left contact and release it to the right contact providing fast accurate electron pumping \cite{Kouw91}. Similar double modulated barrier pumps at GHz frequencies lead to average pump current of several hundred picaoamperes \cite{Blum07} more suitable for metrology. Single gate pumps have been implemented using silicon nanowires \cite{Fuji08} and etched GaAl/GaAsAl nanowires. Recent improvements gave accuracy better than $1.2.10^{-6}$ for a 150pA generated current \cite{Gibl12} making these devices promising for current standard in quantum metrology while similar devices combined with quantum Hall effect may give a short path to metrology triangle closure \cite{Hoh12}.

Another approach to single charge sources is based on moving quantum dots carrying integer charges using of Surface Acoustic Waves (SAWs) \cite{Taly97}. In a piezo-electric material like GaAs, a mechanical wave can be generated by the electric field created by interdigitated gates deposited on top of the device. An electrical potential well moving at the surface phonon velocity, a few $km/s$, can trap electrons confined in a 1D wire and the moving quantum dot so realized can transfer them through the conductor at GHz frequencies leading to good current quantization accuracy \cite{Jans00}.

\subsection{single electron sources for electron quantum optics}
In all the examples cited above, the goal was a good accuracy in the charge transfer. In the following we will focus on quantum experiments appropriate for electron quantum optics where the emitted single electrons play the role of  flying charge qubits \cite{Bert00}. Depending on the system used the charge or the spin may code a binary qubit information. The flying qubit approach is fundamentally different from the standard qubit realization which are based on a static localized two-level system (nuclear or electron spin, atom, charge state of a superconducting metallic island, photonic state of resonator circuits, etc.). Here the information is coded by the presence or not of a particle (electron, photon) propagating or delocalized in spatial modes.

 Experiments in this direction using the SAW technique described above have been done recently. SAW assisted transfer of single electrons through depleted 1D channels have demonstrated single electron transfer over several $\mu m$ between two distant quantum dots \cite{Herm11,McNe11} and the spin state of one and two electrons have been transferred \cite{Bert16}. Using quantum dot sources, single electrons have been emitted at high energy above the conductor Fermi level. The energy and time properties of the wavepacket have been analyzed \cite{Flet13}. The partitioning of spin entangled electron pairs emitted that way has been recently realized \cite{Ubbe14}. These examples are promising for further quantum experiments where only the spin coherence of electrons is concerned, but regarding orbital quantum coherence, appropriate for charge flying qubits, different approaches are needed.

\begin{figure}[h]%
\includegraphics*[width=\linewidth,height=\linewidth]{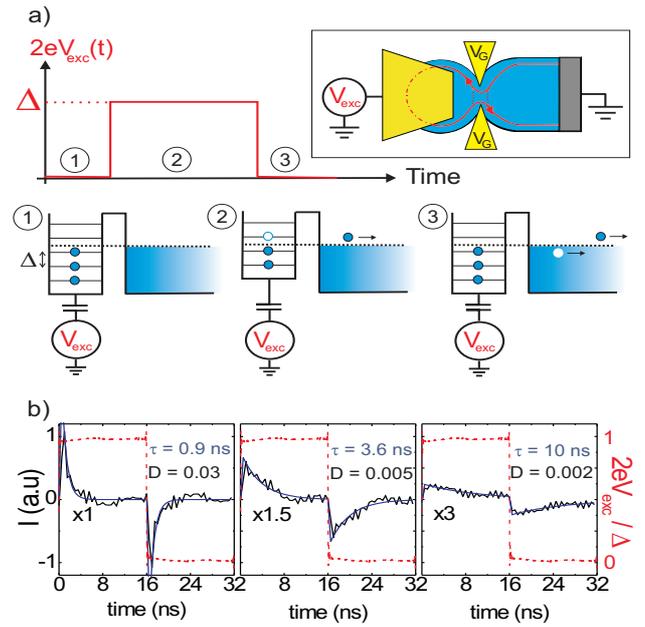}
\caption{%
  a) Radio-frequency pulse applied on the top gate (inset of the figure). (1) Starting point: the fermi level lies between two discrete energy levels of the quantum dot. (2) 2e$V_{exc}(t)$ is equal to the level spacing $\Delta$. An electron escapes the dot at a well defined and tunable energy. (3) $V_{exc}(t)$ is brought back to its initial value, a hole escape at energy below the lead Fermi energy . b) Time domain measurement of the average current (black curves) on one period of the excitation signal (red
curves) at 2e$V_{exc}(t)$ = $\Delta$ for three values of the transmission $D$. The expected exponential relaxation with time $\hbar/D\Delta$ fits well the data (blue curve)(figure adapted from \cite{Feve07}).}
\label{SES}
\end{figure}

 The first coherent single electron source has been proposed and realized by one of the present authors with the ENS Paris team \cite{Feve07}. It is based on a mesocopic capacitor which has been precedently realized \cite{Gabe06} to check the universal quantization of the charge relaxation resistance (called B\"{u}ttiker's resistance $h/2e^{2})$ predicted by M. B\"{u}ttiker \cite{Butt93}. Here no dc current is produced but only an ac current made from the periodic injection of single electrons followed by single holes above and below the Fermi energy $E_{F}$ of the target quantum conductor. For ease of operation and further use in electron quantum optics, the conductor is in the integer quantum Hall effect regime where a strong magnetic field quantizes the electron cyclotron orbits and only 1D chiral modes propagating along the sample edges, called edge channels, ensure electron conduction. Here, the capacitor is made so small that energy level quantization makes it behave as a quantum dot. To ensure energy level and charge quantization, the capacitor is weakly connected to the leads, the chiral edge channels, via a quantum point contact which controls the tunnel coupling. The operating principle is as follows, see Fig.\ref{SES}. Starting from a situation where the last occupied energy level is below the Fermi energy, a sudden rise of the voltage applied on a top gate capacitively coupled to the mesoscopic capacitor rises the occupied energy level above the Fermi energy. After a time equal to the life time $\simeq\hbar/D\Delta$ of the energy level and controlled by the barrier transmission $D$, an electron is emitted at a tunable energy $\varepsilon_{e}$ above the Fermi level ($\Delta$ is the energy level spacing). Then restoring the top gate voltage to its initial value pulls down the energy level below the Fermi energy: an electron is captured or equivalently a hole is emitted with a definite energy $-\varepsilon_{h}$ below the lead Fermi energy.

The mesoscopic capacitor single electron source has been used for electron quantum optics experiments like the partitioning of
single electrons in an electronic beam splitter \cite{Bocq12}, an Hanbury Brown-Twiss experimental set-up commonly used in optics, and the Hong Ou mandel (HOM) electron interferometry where two electron emitted with a controlled relative time delay are sent to an electronic beam-splitter and second order interference revealing particle indistinguishability is observed in the detection statistics at the beam-splitter output \cite{Bocq13}. Similar experiments have been repeated with the leviton source discussed below \cite{Dubo13,Jull14}. The 2007 first realization of a single electron source has triggered many theoretical works \cite{Olkh08,Sple08,Keel08,Sple09,Math10,Math11,Haac11,Jonc12,Batt13} as it opened the possibility to test dynamical quantum transport at the single electron level, investigating single and two-electron coherence, non-local entanglement, waiting time statistics or quantum heat fluctuations. However the mesocopic capacitor source is delicate to operate and to fabricate (tuning two identical energy levels and emission time to demonstrate electron undistinguishability in \cite{Bocq13} was a tour de force). Like single photon sources it displays a fundamental quantum uncertainty in emission time, a quantum jitter measured by high frequency shot noise in \cite{Parm12}. A simpler complementary approach to single electron sources for electron quantum optics came more recently with the realization of voltage pulse sources.

The voltage pulse source is based on a simple principle. A voltage pulse $V(t)$ is applied on the contact of a quantum conductor where electrons can propagate in few modes. According to ac quantum transport laws, if all other contacts are grounded, a current $I(t)=e^{2}/hV(t)$ is injected form the contact in each mode emitted from this contact (disregarding spin).
Then, by tuning the amplitude and duration of the voltage pulse such that $\int_{-\infty}^{\infty}I(t)dt=e$ a single charge is injected. The procedure seems too simple to think it can give correct results. Indeed, if a single charge is actually injected, it is in general accompanied by unwanted neutral excitations in the form of electron-hole pairs resulting from the perturbation of the electrons already present in the conductor (the Fermi Sea). However, nearly twenty years ago, in their quest for a clean electron generation model for studying the Full Counting Statistics in quantum conductors, L. Levitov and collaborators discovered that, if the voltage pulse has the shape of a Lorentzian, a clean single electron injection (without extra neutral excitations) can be realized \cite{Levi97,Ivan97,lebe05,Keel06}. The Lorentzian voltage pulse single electron source has been experimentally realized in 2013 by the present authors and their team at CEA Paris-Saclay and the predicted minimal excitations states, now called levitons, have been carefully characterized by their minimal shot noise \cite{Dubo13}. The single electron leviton source is presented in more detailed and reviewed in the next parts of the present paper.

\begin{figure}[t]%
\includegraphics*[width=\linewidth,height=\linewidth]{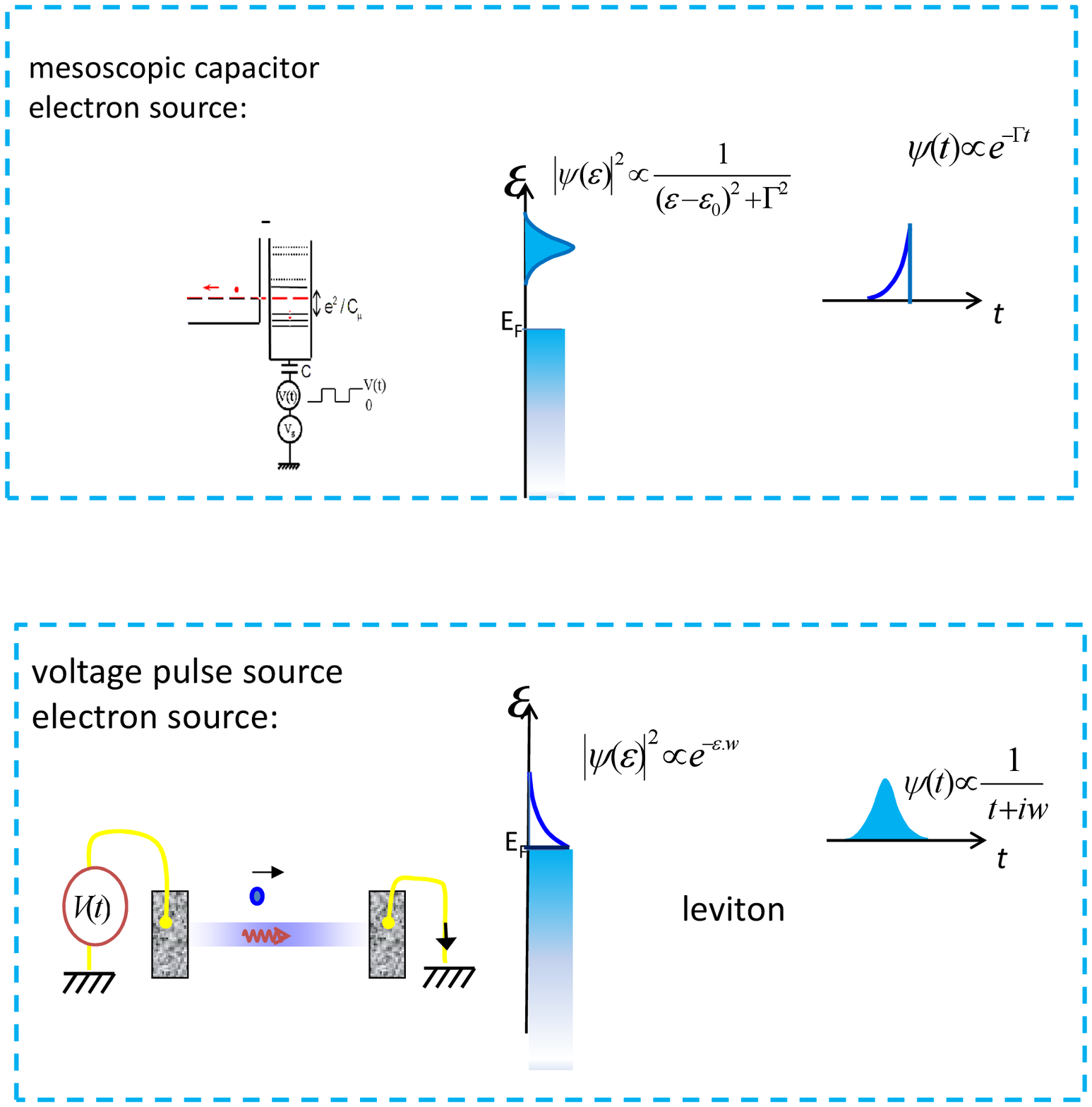}
\caption{%
  Top figure: the mesoscopic capacitor source (energy-resolved source). Its energy distribution is lorentzian and the time wavepacket envelope is semi-exponential. Bottom Figure: the voltage pulse leviton source. Its energy distribution is semi exponential and the time wavepacket is lorentzian. The two sources are suitable for electron quantum optics. They show a duality when interchanging time and energy }
\label{SESLev}
\end{figure}

In order to conclude this part I will refer the reader to Fig.~\ref{SESLev} which shows a comparison of the Mesoscopic capacitor and Leviton single electron sources, both being suitable to perform electron quantum optics. The figure shows the complementary aspects of the two sources. The mesoscopic capacitor source emits energy resolved electrons while the leviton source is time-resolved. There is a duality of their properties if we interchange time and energy: Lorentzian energy distribution versus Lorentzian time wavepacket, semi-exponential time wavepacket versus semi exponential energy distribution above the Fermi energy. Being energy-resolved, the mesoscopic capacitor source allows to make nice fundamental tests of the energy relaxation and decoherence of a Landau like quasiparticle above the Fermi sea \cite{Ferr14}. For the same reason the electrons injected well above the Fermi sea are more prone to decoherence than the levitons whose energy distribution is as close as possible to the Fermi energy \cite{Ferr14}. Due to charging effects complicating electron injection, the mesoscopic capacitor source is limited to single charge injection. The voltage pulse source lacks of exact charge quantization (as the charge is tunable) but it can simultaneously injects any number of electrons while keeping the levitonic minimal excitation property. Interestingly, it was theoretically shown that if the mesoscopic capacitor is driven adiabatically (i.e. the energy level not suddenly risen above the Fermi energy but slowly risen at constant speed), the Lorentzian energy shape of the level gives rise to a Lorentzian wavepacket similar to a leviton with  the emitted charge fundamentally fixed to the charge e \cite{Keel08}.

\section{Levitons: time-resolved single electron with minimal excitation property.}
\subsection{principle}
We consider a single mode conductor, spin disregarded. The following results can be directly extended to multiple modes including spin. The single mode can be an edge channel of the integer quantum Hall effect for filling factor one. It can also be a single electronic mode of a 2D conductor spatially filtered by a narrow constriction like a Quantum Point Contact inserted in the middle of the 2D conductor.
When a time dependent voltage $V(t)$ is applied, say on the left contact of conductor while the opposite contact is grounded, electrons emitted at energy $\varepsilon$ below the Fermi level of the contact, and experiencing the voltage, acquire a time dependent phase $\phi(t)=e\int_{-\infty}^{t}V(t')dt'/\hbar$. Because of the time dependency, energy is not conserved and the electron are scattered in a superposition of quantum states of different possible energies. The amplitude of probability to have the energy displaced from $\varepsilon$ to $\varepsilon+\delta\varepsilon$ is :
\begin{equation}
\label{eqp}
p(\delta\varepsilon)= \int_{-\infty}^{\infty} e^{-i\phi(t)} e^{i\delta\varepsilon t/\hbar} dt
\end{equation}
and the probability $P(\delta\varepsilon)=|p(\delta\varepsilon)|^{2}$. Eq.~\ref{eqp} is the basis of the Floquet scattering theory developed by M. B\"{u}ttiker and M. Moskalets for periodically driven conductors \cite{Moska02}. For $V(t)$ a voltage pulse, the injected charge $q$ is
$q=\int_{-\infty}^{\infty}I(t)dt$. For a spinless single mode, for which $I(t)=e^{2}V(t)/h$, this gives the Faraday flux (or action):
\begin{equation}
\label{action}
\int_{-\infty}^{\infty}eV(t)dt = (q/e)h
\end{equation}
and the phase increment:
\begin{equation}
\label{phaseincr}
\Delta \phi = \phi(-\infty)-\phi(\infty)=2\pi (q/e)
\end{equation}
In general $p(\delta\varepsilon)$ have non zero values for both positive and negative $\delta\varepsilon$. Consequently all electrons of the left contact Fermi sea are displaced up and down in energy. Compared with the reference Fermi sea of the grounded right contact, this gives both electron and hole like excitations. At zero temperature, with sharp Fermi sea distribution, one can quantify the number of electrons $N_{e}$ and hole $N_{h}$ excitations created by the pulse. They are given by:
\begin{equation}
\label{Ne}
N_{e}=\int_{0}^{\infty}\delta\varepsilon P(\delta\varepsilon)d(\delta\varepsilon)
\end{equation}
\begin{equation}
\label{Ne}
N_{h}=\int_{-\infty}^{0}(-\delta\varepsilon)P(\delta\varepsilon)d(\delta\varepsilon)
\end{equation}
The charge introduced by the pulse in the conductor is $q=e(N_{e}-N_{h})$ and the total number of excitations is $N_{exc}=N_{e}+N_{h}$.

\begin{figure}[t]%
\includegraphics*[width=\linewidth,height=\linewidth]{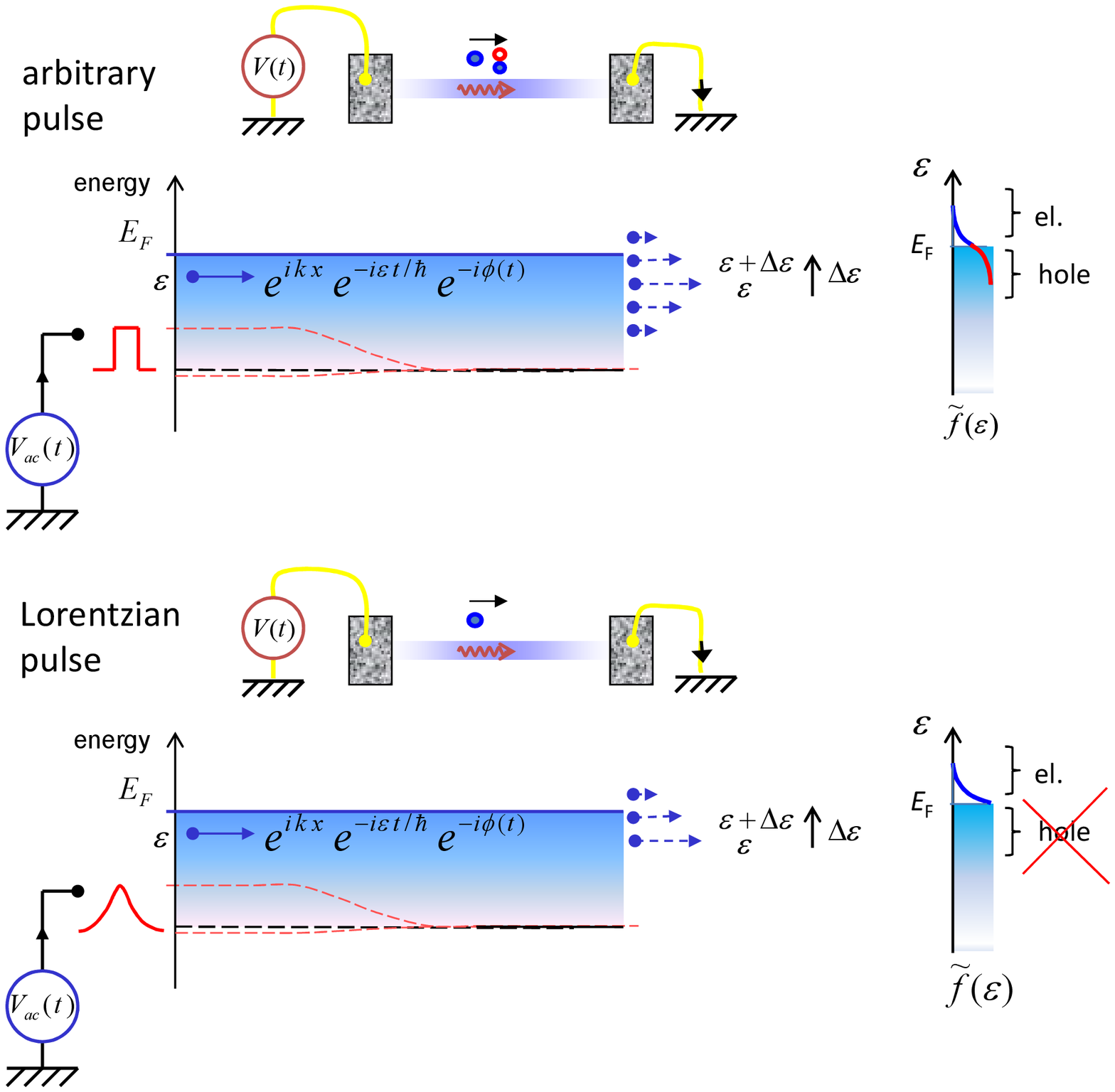}
\caption{%
  Top figure: effect of an arbitrary pulse on the Fermi sea. Electrons are scattered to different energies above and below their initial energy. Electron hole excitations are created, as schematically represented in the energy distribution $\widetilde{f}(\varepsilon)$. Bottom figure: applying a Lorentzian shape voltage pulse generates a minimal excitation state called Leviton , no hole excitation is created.}
\label{ArbiPulse}
\end{figure}

For a single electron source injecting electrons in the form of a minimal state called leviton, it is mandatory to have no hole excitations creation, i.e. $N_{exc}=N_{e}=1$ such that only excitation is the injected electron, as schematically shown in Fig~\ref{ArbiPulse}. To do that, the Fourier transform of the phase term $e^{-i\phi(t)}$, which in Eq.~\ref{eqp} gives $p(\delta\varepsilon)$, must be zero for all negative $\delta\varepsilon$. According to standard complex integration technique, this implies the phase term $e^{-i\phi(t)}$, prolongated in the complex plane must have no pole in the lower half plane and at least one pole in the upper half complex plane. The simplest expression for the phase term with only one pole, at say $t=iw$, is:
\begin{equation}
\label{onepole}
e^{-i\phi(t)}=\frac{t+iw}{t-iw}
\end{equation}
With this form, the phase increment is $\Delta \phi=2\pi$ and, from Eq.\ref{phaseincr}, the associated charge is $q=e$ while the phase derivative is a Lorentzian with $2w$ full width at mid-height. The corresponding voltage pulse is therefore a Lorentzian:
\begin{equation}
\label{VLor}
V(t)=\frac{\hbar}{e}\frac{2w}{t^{2}+w^{2}}
\end{equation}

Eqs. \ref{onepole},\ref{VLor} uniquely define the type of voltage pulse needed to generate a single charge leviton. Generalization to multiple electron injection in the form of levitons injected at time $t_{k}$ with width $w_{k}$ is possible by introducing extra poles in the upper complex plane, such that
\begin{equation}
\label{multipole}
e^{-i\phi(t)}=\prod_{k=1}^{N}\frac{t-t_{k}+iw_{k}}{t-t_{k}-iw_{k}}
\end{equation}
\begin{equation}
\label{VLorMulti}
V(t)=\sum_{k=1}^{N}\frac{\hbar}{e}\frac{2w_{k}}{(t-t_{k})^{2}+w_{k}^{2}}
\end{equation}
The absence of poles in the lower half part of the complex plane in \ref{multipole}, ensures that the multiple electron injection still forms a minimal excitation state.
Simultaneous injection of N electrons in the form of levitons is also possible by simply increasing by $N$ the pulse amplitude. The phase term acquires a single pole of order N, in the upper half complex plane $e^{-i\phi_{N}(t)}=(\frac{t+iw}{t-iw})^{N}$. In \cite{Glat16} it was remarked that such phase modulation generates $N$ orthogonal envelope wave-functions of the form $\phi_{k}(u)=\frac{(u+iw)^{k-1}}{(u-iw)^{k}}$, with $u = t-x/v{F}$, $k=1,N$ and the resulting quantum state generated from the voltage pulsed Fermi sea is a slater determinant made of these $N$ wave-functions.

The voltage pulse charge injection method also allows to inject arbitrarily a non-integer charge $q$. Keeping a Lorentzian pulse shape,
this gives the phase term:
\begin{equation}
\label{fracpole}
e^{-i\phi(t)}=\bigg( \frac{t+iw}{t-iw}\bigg)^{q/e}
\end{equation}
One immediately sees that if $q$ is not integer, the resulting fractional power gives non-analytic properties to the phase term on
the real axis. As a consequence, the Fourier transform of $e^{-i\phi(t)}$ acquires non-zero values for positive and negative $\delta\varepsilon$ giving both electron and hole excitations. This situation was named in Levitov's original paper \cite{Levi97} "dynamical orthogonality catastrophe" in reference to the Anderson orthogonality catastrophe problem but in the time domain.

To summarize, Lorentzian voltage pulses generate clean minimal excitation states called levitons carrying an integer number of charge only. Other pulse shape introduce extra excitations in the form of neutral electron hole pairs.

\subsection{Evidence and characterization of levitons}
\subsubsection{Minimal noise for minimal excitation states}
In a practical experiment, as in \cite{Dubo13}, periodic injection is chosen. For an injection frequency $\nu$, period $T=1/\nu$, the average current is $\overline{I}=e\nu$. The periodicity implies that $P(\delta\varepsilon)=\Sigma_{l=-\infty}^{\infty}P_{l}\delta(\delta\varepsilon-lh\nu)$, the discrete $P_{l}$ being the $l$-photo-absorption probabilities.  To experimentally demonstrate that the right pulse shape, the Lorentzian, generates a minimal excitation state, the expected levitons were sent to an artificial quantum impurity, a Quantum Point Contact, which partially transmits a single electronic mode. Its mode transmission $D$ was tunable by gates and measurable by dc conductance. The idea is to take advantage of the partitioning of electron and hole excitations by the QPC into transmitted and reflected modes. This fundamentally quantum probabilistic effect generates what is known as quantum shot noise,where shot noise means current noise. It can be shown that the shot noise at zero temperature is a direct measure of the total number of excitations $N_{exc}=N_{e}+N_{h}$ \cite{Levi97,Keel06,Dubo12}. The spectral density of the current noise $S_{I}$ is given by:
\begin{equation}
\label{ShotN}
S_{I}=2e^{2}\nu D(1-D)(N_{e}+N_{h})
\end{equation}

By comparing the noise generated with sinewave, square wave and Lorentzian pulses, ref.\cite{Dubo12} showed that for integer charge, thermal corrections included, the shot noise of Lorentzian pulses is indeed minimal and corresponds exactly to the noise of a clean single charge. Thus minimal excitation states, levitons, generate minimal noise.

\subsubsection{energy spectroscopy: absence of hole excitations}
Energy spectroscopy is another way to characterize the levitonic state. Indeed, contrary to sinewave or square wave pulses, it is expected that Lorentzian pulses carrying charge e generate no hole excitation. It is thus important to show the absence of hole excitations for energies below the Fermi energy. Again, the partition shot noise of a Quantum Point Contact can be used to demonstrate this. The idea is to add a dc voltage $V_{R}$ on the opposite right contact. Under negative bias, the hole emitted from the periodically driven left contact in the energy range $-eV_{R}<\varepsilon<0$ will no longer contribute to noise and the corresponding shot noise variation will measure their number. For positive $V_{R}$, electrons emitted in the energy range $0>\varepsilon>V_{R}$ will anti-bunch with electron excitations coming from the driven left contact and no noise is produced. Again, the noise variation with $V_{R}$ will give a measure of the electron excitations. With a finite $V_{R}$, the shot noise is still given by expression Eq.~\ref{ShotN} but for $V_{R}>0$, $N_{e}$ is replaced by $N_{e,V_{R}}=\int_{EV_{R}}^{\infty}\delta\varepsilon P(\delta\varepsilon)d(\delta\varepsilon)$ and $N_{h}$ unchanged while for $V_{R}<0$, $N_{e}$ is unchanged by $N_{h}$ relaced by $N_{h,V_{R}}=\int_{-\infty}^{-e|V_{R}|}(-\delta\varepsilon)P(\delta\varepsilon)d(\delta\varepsilon)$.

\begin{figure}[t]%
\includegraphics*[width=\linewidth,height=\linewidth]{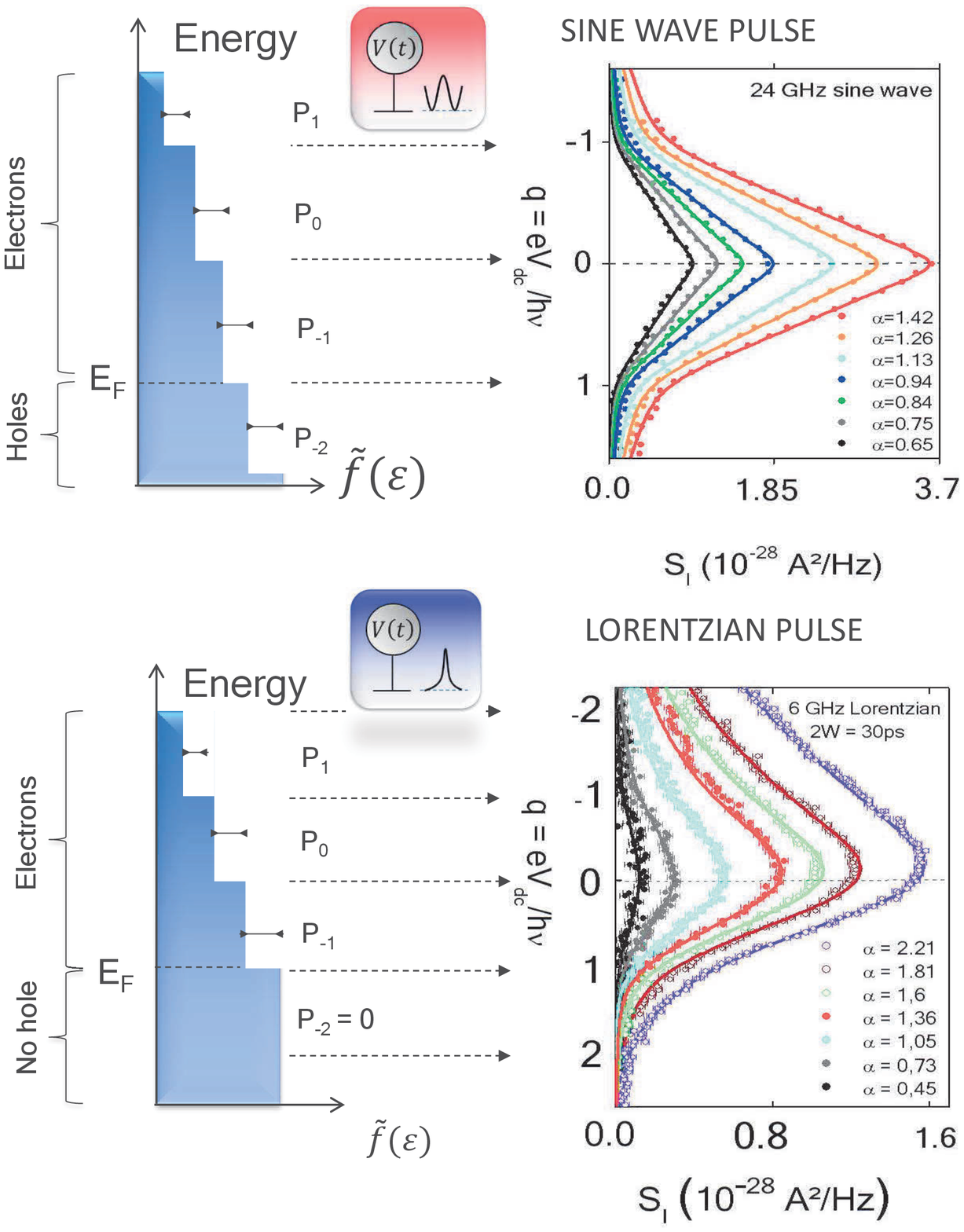}
\caption{%
  Top figure: shot noise spectroscopy of a sine wave pulse. Since an equal number of electron and hole excitations is created the noise spectroscopy is symmetric with respect to $V_R$ = 0.The schematic zero temperature energy distribution $\widetilde{f}(\varepsilon)$ is shown on the left. Bottom figure: shot noise spectroscopy of a lorentzian pulse. The number of holes is minimized to zero and the shot noise spectroscopy is asymmetric. The figures are adapted from \cite{Dubo13}}
\label{Spectro}
\end{figure}

Fig.~\ref{Spectro} shows shot noise spectroscopy comparison of sine wave and Lorenztian pulses and comparison with predictions. For a sine wave, a symmetric noise variation with $V_{R}$ is observed due to symmetric electron and hole contribution which strikingly differs from the asymmetric noise variation for Lorentzian due to the absence of holes.

\subsubsection{Time domain characterization: Hong Ou Mandel correlations }
In the celebrated Hong Ou Mandel photonic experiment, the authors were at first interested to determine the time shape of the photon pairs emitted by parametric down conversion of a Laser pulse sent to a non-linear crystal. To do this, they took advantage of the photon bunching effect which is expected to occur when two undistinguishable photons arrive at the same time and mix on a beam-splitter. According to Bose statistics, both photons prefer to exit at the same beam-splitter output and experimentalists repeating the experiment will find fluctuations $\langle(\Delta N)^{2}\rangle$ in the detection photon number $N$ due to the partitioning by the beam-splitter of two bunched photons. This is twice the particle noise expected if the photons were separately arriving and partitioned by the beam-splitter. By varying the delay $\tau$ between the photon arrival, one progressively goes from the first to the second situation and the variation is simply given by the overlap of the photon wave-functions $\psi$ which partially mix for finite $\tau$ in the beam-splitter, i.e. $\langle(\Delta N)^{2}\rangle_{\tau}=\langle(\Delta N)^{2}\rangle_{\tau=\infty}(1+|\langle\psi(0)|\psi(\tau)\rangle|^2)$. This provided a measure of the photon wave-packet extension and a nice evidence of bosonic quantum statistics. A similar trick can be done with electrons which, being fermions, anti-bunch. When in coincidence ($\tau=0$) they will always take different outputs to satisfy Pauli exclusion and the particle noise is zero.

Hong Ou Mandel (HOM) correlations can be realized using levitons periodically sent with a relative delay $\tau$ from opposite contacts toward a Quantum Point Contact playing the role of the electronic beam-splitter while measuring the cross correlation shot noise between output leads. One expects
\begin{equation}
\label{ShotHOM}
S_{I}=2e^{2}\nu D(1-D)2(1-|\langle\psi(x)|\psi(x-v_{F}\tau)\rangle|^2)
\end{equation}
where $\psi$ is the leviton wavefunction. If doubly charge levitons are sent, one expects:
\begin{equation}
\begin{split}
\label{ShotHOM2}
S_{I}=2e^{2}\nu D(1-D)2(2-|\langle\psi_{1}(x)|\psi_{1}(x-v_{F}\tau)\rangle|^2-\\
|\langle\psi_{2}(x)|\psi_{2}(x-v_{F}\tau)\rangle|^2)
\end{split}
\end{equation}
where $\psi_{1,2}$ are the first two orthogonal levitonic wave functions \cite{Glat16} carrying the two incoming electrons. Remarkably, the voltage pulse electron injection technique allows the generation of an arbitrary number of electrons and for performing N-electrons HOM correlation. In this case,  Eq.~\ref{ShotHOM2} generalizes easily. Because of constraints due to Fermi statistics, a direct comparison between HOM correlations with N-photon Fock states \cite{Lalo12} and those with N-electron minimal excitation states cannot be done. This situation opens new perspectives in quantum mechanics.

\begin{figure}[t]%
\includegraphics*[width=\linewidth,height=\linewidth]{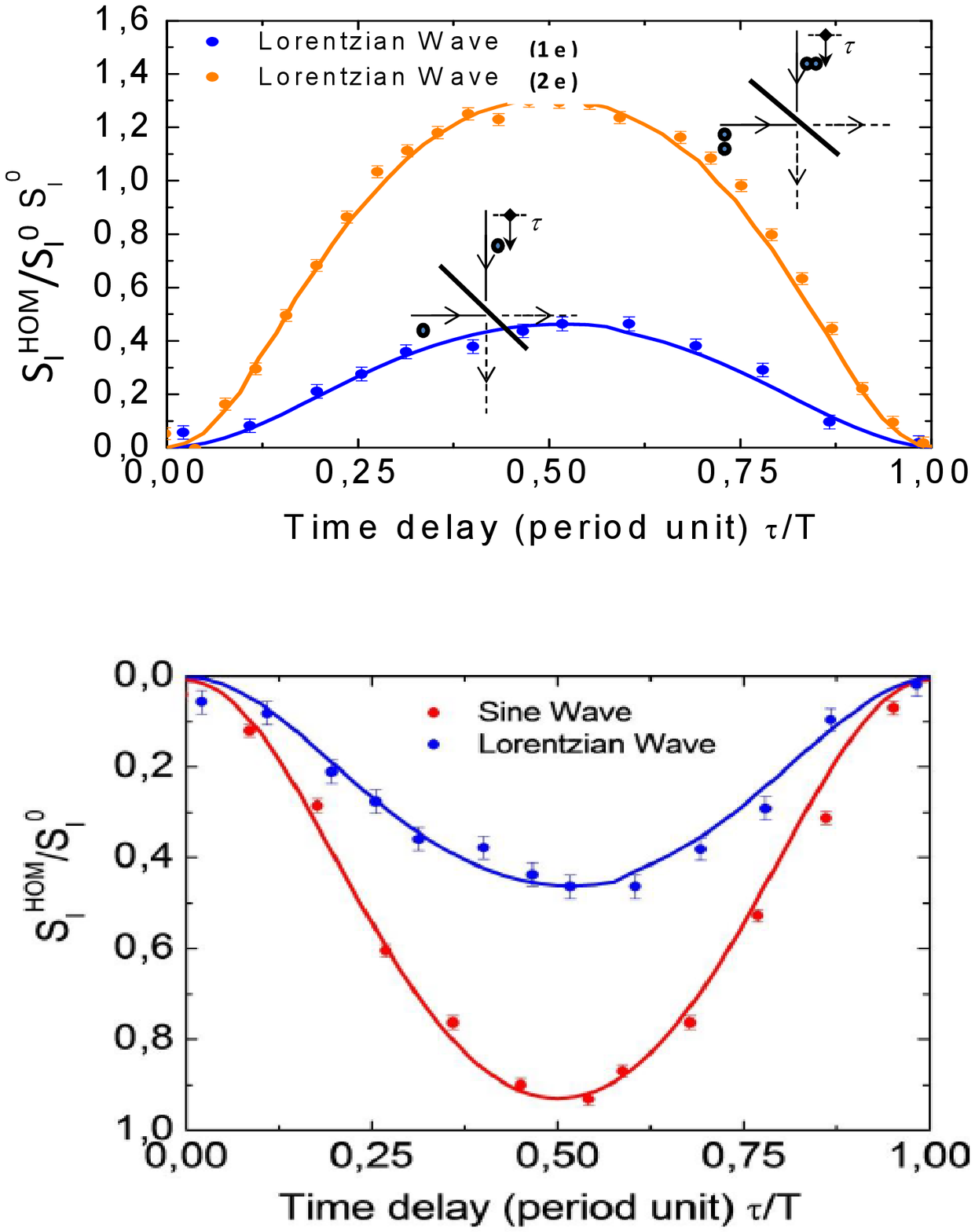}
\caption{%
  Top figure: HOM experiment for one and two electrons injected in the separate input of a beam-splitter. The HOM shot noise resulting from partitioning of antibunched electrons is plotted versus the arrival  time delay $\tau$ giving access to the wavefunction overlaps. Note the excellent agreement with the expected theory (solid line). Bottom figure: HOM experiment with two different pulses.  The sine pulse creating a large amount of electron hole excitations, the HOM shot noise is enhanced. The figures are adapted from \cite{Dubo12}.}
\label{HOM}
\end{figure}

Fig.~\ref{HOM} shows the HOM shot noise experiment done with single and doubly charged levitons. The comparison with expected HOM noise variation is excellent without adjustable parameters. This provides a good check of the levitonic states with one or two electrons in the time domain. The HOM shot noise of colliding sine-wave pulses is also shown for comparison. Here, because sine-wave pulses give rise to charge pulses accompanied with a cloud of electron-hole pairs and all these excitations interfere in the electron beam-splitter, the physical information to extract is not clear (except to confirm the good agreement with HOM shot noise theoretical modeling \cite{Dubo12}).

The leviton HOM correlations give remarkably 100\% noise suppression at $\tau=0$. This contrasts with the small HOM dip observed in ref \cite{Bocq13} where electron are injected from a mesocopic capacitor. In the latter case, the reason for a weak HOM noise suppression is not due to the different nature of the source but is due to the propagation in the leads between the source and the beam-splitter. In \cite{Bocq13} the leads are made of two chiral edge channels (Integer Quantum Hall regime at filling factor 2) and electrons are injected in the outer edge channel. A (pseudo)-spin charge separation occurs where the incoming electron fractionalizes. This induces decoherence and energy relaxation which spoils the otherwise naively expected full HOM noise suppression. Working with a single edge or putting the source closer to the beam-splitter would produce a better HOM shot noise dip.

\begin{figure}[t]%
\includegraphics*[width=\linewidth,height=\linewidth]{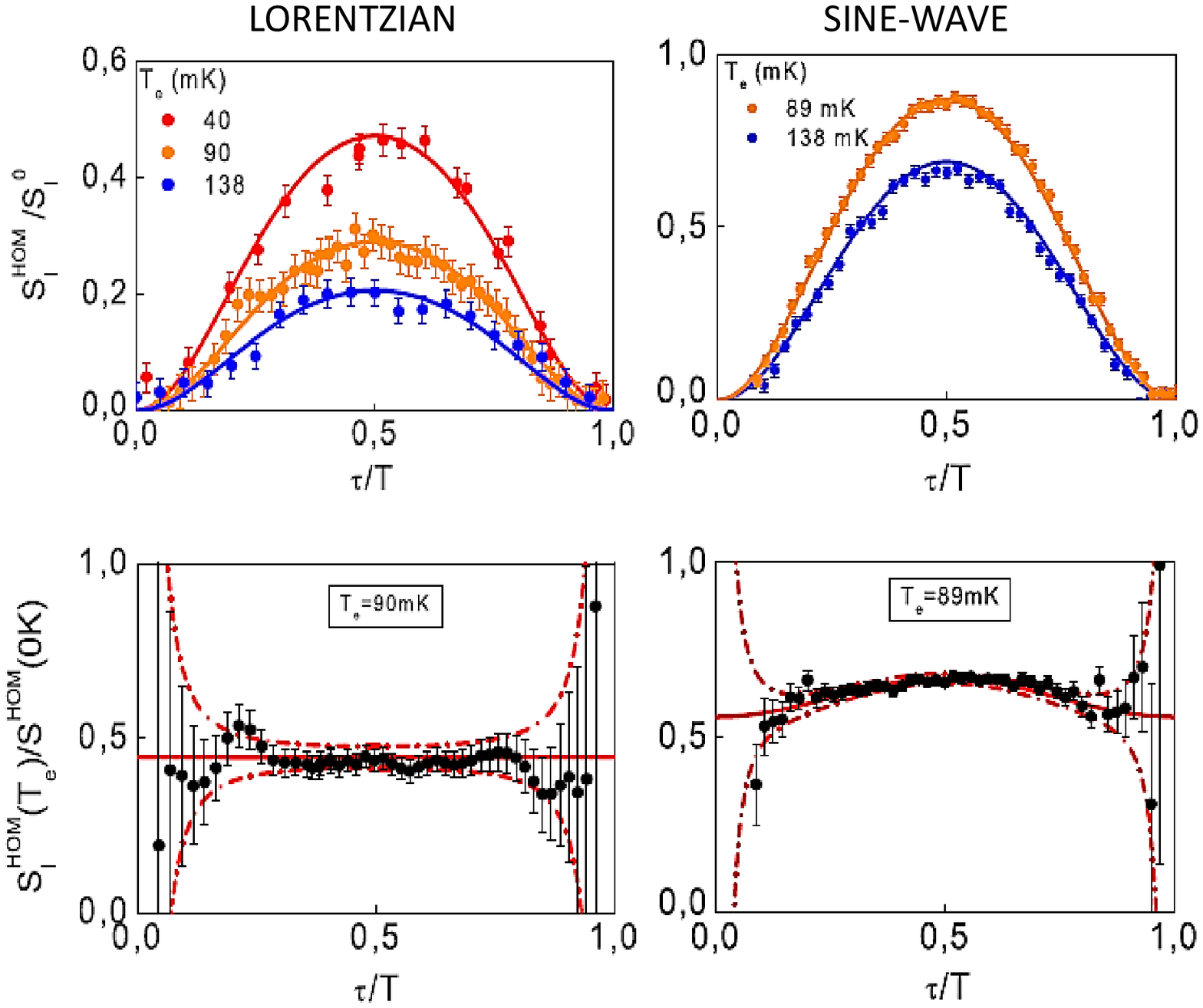}
\caption{%
  Temperature dependence of the HOM shot noise for lorentzian (left) and sinewave (right) pulses. By plotting the ratio of the HOM noise  to the theoretical zero temperature HOM noise versus the HOM delay $\tau$, one observes that, contrary to the sinewave pulses, for levitons the HOM noise shape versus $\tau$ is not affected by thermal excitations. Figures adapted from \cite{Glat16}}
\label{HOMtempe}
\end{figure}

In \cite{Dubo12}, it was remarked that the HOM noise shape versus $\tau$ does not depends on temperature. This was actually experimentally observed \cite{Glat16}. This remarkable decoupling occurs only for single charge levitons but not for sine wave pulses nor for doubly charged  levitons. The robustness of the 100\% HOM dip to temperature for any voltage pulse shape and of the HOM leviton noise curve versus $\tau$ to temperature effects and decoherence has been recently theoretically studied in \cite{Mosk16}. Fig~\ref{HOMtempe} shows experimental HOM noise measurements demonstrating the decoupling between the temperature and the delay $\tau$ for the case of a leviton and shows the absence of decoupling for sinewave pulses.

\subsubsection{Full characterization: Wigner function of a leviton}
In the last two sections levitons have been characterized in time or in energy. But a wave function is a more complex object. Because of the fundamental quantum uncertainty of conjugated variables ( energy versus time or momentum versus position), standard quantum measurements gives in general information on the wavefunction versus one variable only. Having all information on the wave function requires a lot of different measurements to 'see' the wavefunction under all aspects. This is done using Quantum State Tomography (QST). It consists in measuring the energy density matrix $\varrho(\varepsilon',\varepsilon)=<\psi^{\dag}(\varepsilon')|\psi(\varepsilon)>$
or the coherence $<\psi^{\dag}(t')|\psi(t)>$ to which the pure Fermi sea contribution has been subtracted. From this, one can reconstruct the Wigner function $W(\overline{t},\varepsilon)$ which represents a quasiprobability distribution of the state versus the two-dimensional phase space of conjugated variables  $(\varepsilon,t)$ ( or $(p,x)$) .
\begin{equation}
\label{WignerEq}
W(\overline{t},\varepsilon)=\int_{-\infty}^{\infty}d\delta\varrho(\varepsilon+\delta/2,\varepsilon-\delta/2)e^{(-i\delta \overline{t}/\hbar)}
\end{equation}

\begin{figure}[t]%
\includegraphics*[width=\linewidth,height=\linewidth]{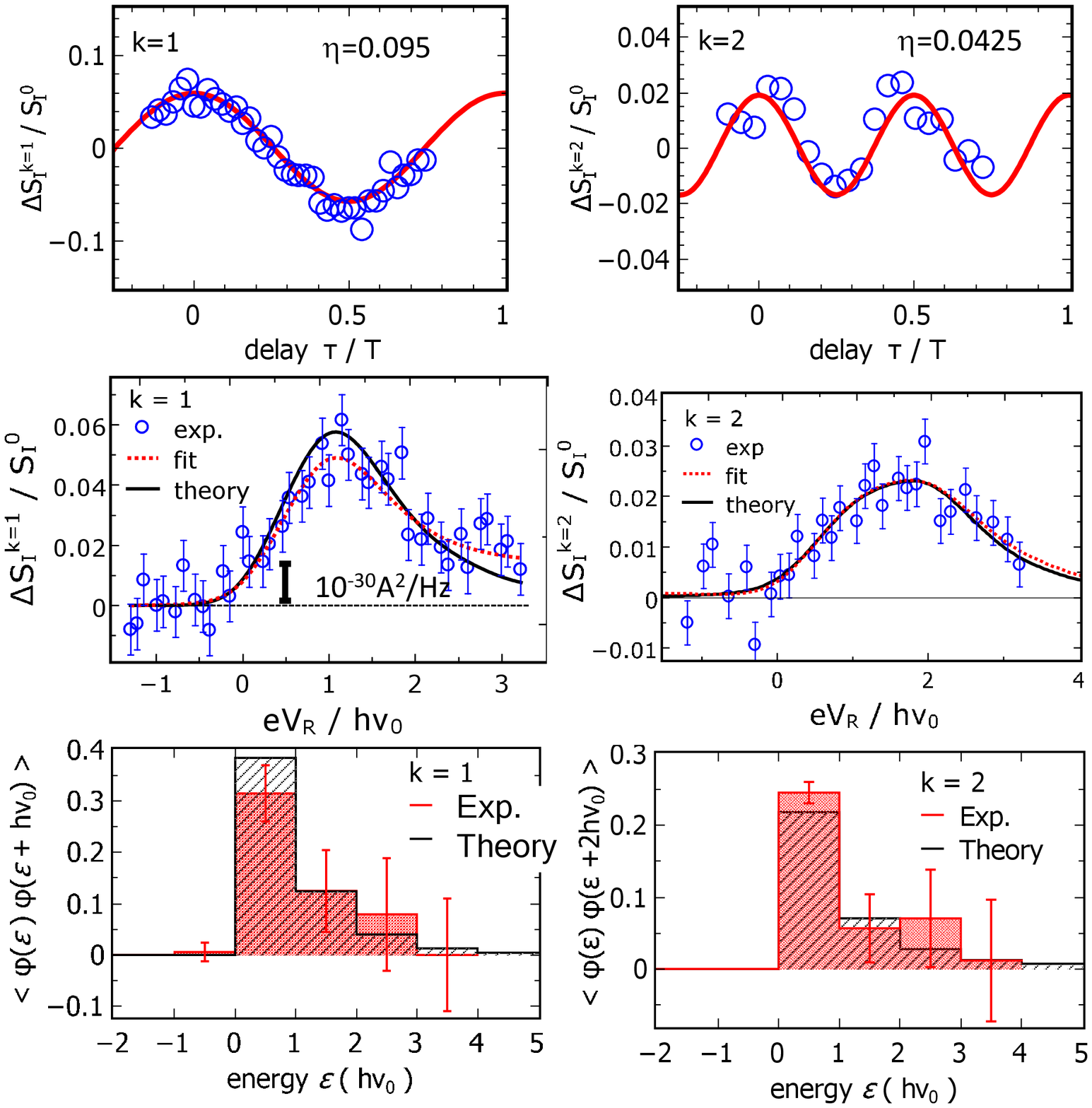}
\caption{%
 The  off diagonal terms of the energy density matrix of a leviton (bottom figures) are extracted from the shot noise  (middle figures) measured when performing the HOM interference of a leviton mixing (colliding) in a beam-splitter with a small sine-wave pulse at frequency $\nu$ (left figure) and $2\nu$ (right figure). The top figures show the noise variation with the time delay $\tau$ between the small sinewave pulse and the periodic levitons. The oscillations, and the doubling of the period at frequency $2\nu$ with respect to $\nu$, prove the existence of non-diagonal coherence terms in the energy density matrix. The figures are adapted from \cite{Jull14}.}
\label{TomoNoise}
\end{figure}

\begin{figure}[t]%
\includegraphics*[width=\linewidth,height=\linewidth]{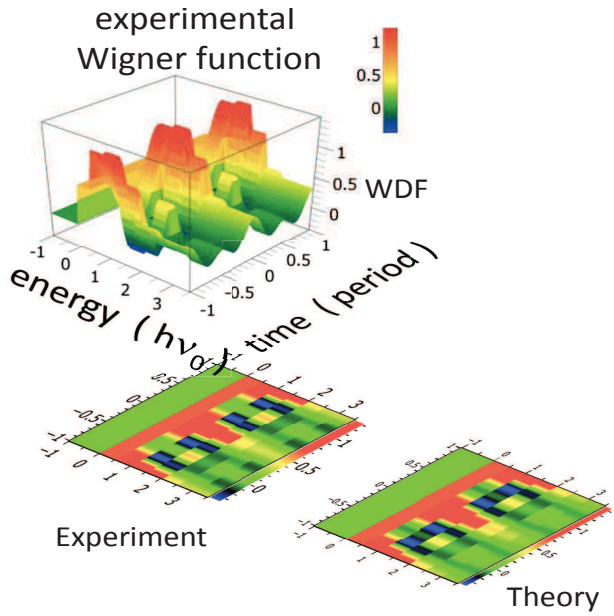}
\caption{%
  Experimental Wigner function extracted from the off-diagonal components of the density matrix (adapted from \cite{Jull14}).}
\label{Wigner}
\end{figure}

For periodic injection, only the non-diagonal elements of $\rho$ differing by a multiple of the fundamental frequency $\varepsilon'-\varepsilon=lh\nu$ are non zero. This implies that the Wigner function is a periodic function of time and this conveniently restricts the number of measurements to be done. A Quantum State Tomography procedure for Measurements of $\varrho(\varepsilon',\varepsilon)$ has been theoretically proposed by C. Grenier in \cite{Gren11} and has been experimentally implemented and reported by the present authors in \cite{Jull14}. First let us remark that the Shot noise spectroscopy makes a measurement of the diagonal term of the energy density matrix as one can deduce from the measurements the electron distribution function $\widetilde{f}(\varepsilon)=\varrho(\varepsilon,\varepsilon)$. In order to access the off-diagonal elements where the energies differ by $kh\nu$, the same shot noise spectroscopy technique is used but, following \cite{Gren11}, instead of only applying a dc voltage $V_{R}$ on the right reservoir, a small amplitude $V_{ac}$ sine-wave voltage is added at frequency $k\nu$. Electrons emitted by the right reservoir at energy $\varepsilon$ are thus is weak amplitude superposition of states of energies $\varepsilon\pm kh\nu$. Mixed in the beam-splitter, they antibunch with the $ \varepsilon$ and $\varepsilon\pm kh\nu$ leviton energy components. Finally, Introducing a delay $\tau $ between the small ac sine-wave signal and the leviton pulses, like in the HOM experiment, provides a way to give a convincing signature  of the interference of the $\varepsilon$ and $\varepsilon+kh\nu$ terms in the resulting shot noise variation. In \cite{Jull14} only the first two harmonic terms were measured ($k=0,1,2$) but the third one could have been measured as well.
The expected shot noise difference $\Delta S_{I}$ obtained when the noise is measured with $V_{ac}$ on and off is given by:
\begin{equation}
\begin{split}
\label{Tomo}
\Delta S_{I} = & S_{I}^{0}2 k \eta \cos(2\pi k\nu\tau) .\\
& .\int_{0}^{eV_{R}}(\varrho(\varepsilon,\varepsilon+kh\nu)-\varrho(\varepsilon,\varepsilon-kh\nu))d\varepsilon/h\nu
\end{split}
\end{equation}
from which the off diagonal terms can be extracted.
Fig~\ref{TomoNoise} shows the noise variations for $k=1,2$. The characteristic doubling of the period with $\tau$ ensures that interference terms are indeed measured. This demonstrates the fully coherent nature of the levitonic quantum state. The deduced off-diagonal components (bottom of the figure) are then used to extract from Eq.~\ref{WignerEq} the leviton Wigner function truncated to the first two harmonic temporal components as shown in Fig.~\ref{Wigner}

\section{Perspectives}
\begin{figure}[t]%
\includegraphics*[width=\linewidth,height=\linewidth]{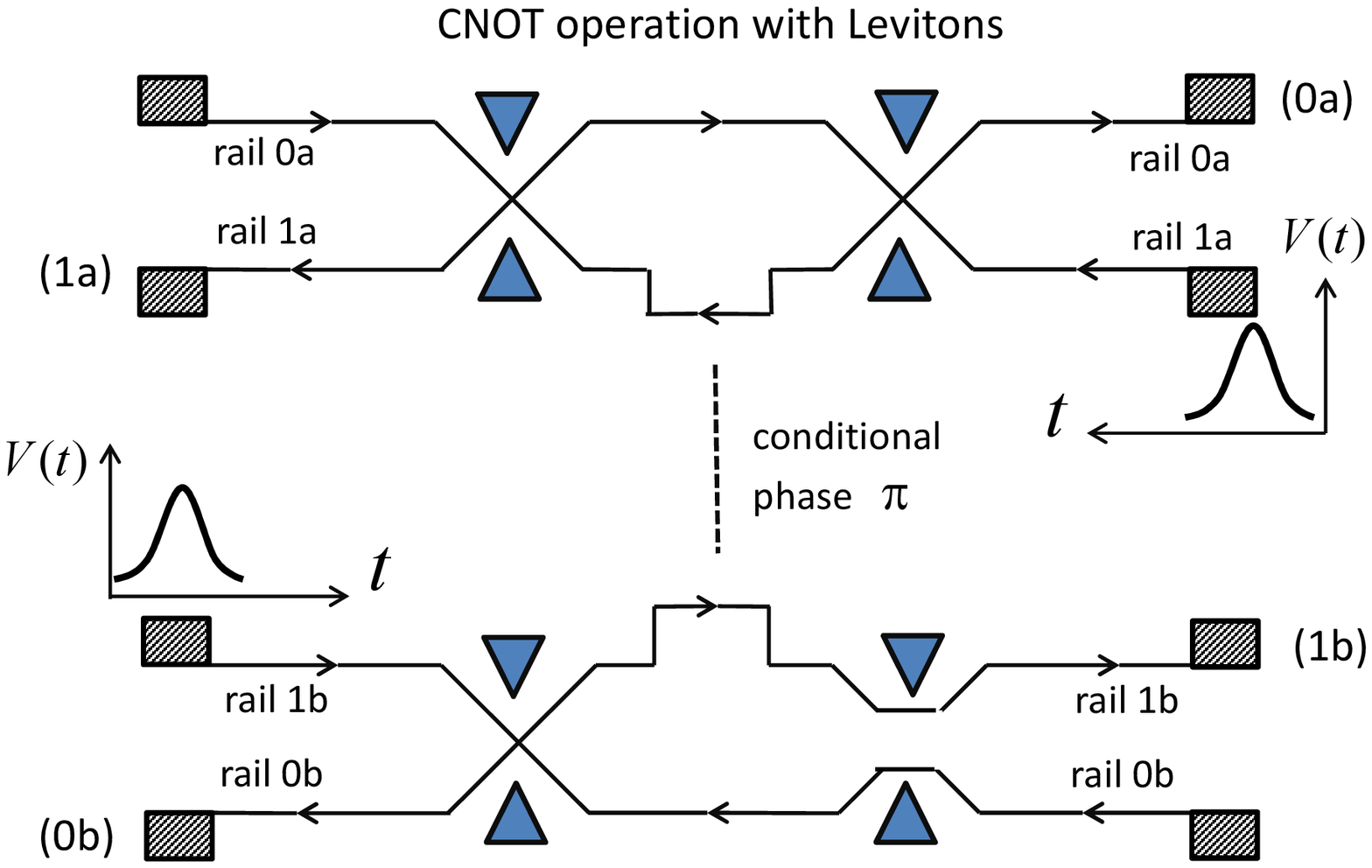}
\caption{%
  Schematic representation of a CNOT quantum gate with Levitons. Two Levitons are injected in rail 1a and 1b. The output state is a maximally entangled state. The vertical dashed line schematically indicates the Coulomb interaction which could provide a conditional $\pi$ phase shift in the two particle (two-qubit) wavefunction.}
\label{CNOT}
\end{figure}

Levitons seem to be the ideal source for electron quantum optics and offers wide perspectives.
The excellent ability to synchronize leviton sources could be exploited to realize  leviton flying qubits using the edge state topology. As an example, we propose the following CNOT quantum gate coupling two flying qubits (a) and (b). Starting with the sate $|1a\rangle|1b\rangle$ by injecting levitons on the appropriate contacts, the output is expected to be $|1a\rangle|1b\rangle + |0a\rangle|0b\rangle$ a Bell state. Here we have used a coulomb coupling which provides a conditional $\pi$ phase shift when an electron on the lower edge of qubit (a)  performs a distant scattering with an electron propagating on the upper edge of qubit (b).
Fig.~\ref{CNOT} schematically displays such a quantum circuit with chiral ballistic electrons to perform the CNOT quantum gate.

Leviton single electron sources have stimulated many theoretical works. The time domain control and the special minimal excitation state property leads to well posed problems not accessible to dc voltage electron sources. Here we will cite only a few examples of many theoretical predictions concerning levitons. For example, the non-local entanglement of electrons emitted by two independent dc voltage sources, as first considered in \cite{Samu04} in a two particle Hanbury Brown Twiss set-up, was difficult to characterize and found very fragile with respect to thermal fluctuations \cite{Samu09}.  The same problem has been recently recently revisited using leviton sources and it has been shown that non local characterization of entanglement is possible without the need of post-selection \cite{Dasen16}. Another example is the generation of neutral levitons pairs by a controlled time variation of the transmission amplitude of a punctual scatter like a QPC which has have been first predicted in \cite{JinZ09}. The principle is based on a controlled variation of the QPC transmission $d(t)$ and reflection $r(t)$ amplitudes. Writing the scattering amplitudes in the form of $d(t)=\sin\phi(t)$ and $r(t)=i \cos(\phi(t)$, to ensures unitarity of the QPC scattering matrix, and choosing a phase variation $\phi(t)$  corresponding to that used to produce levitons with voltage pulses, see Eq.~\ref{onepole}, neutral leviton anti-leviton pairs are predicted to be emitted in the adjacent leads from the modulated scatter. These pairs are expected to be entangled \cite{Sherk09}. Embedding, these neutral leviton pairs in a Mach-Zehnder electronic interferometer, their entanglement has been theoretically demonstrated and the entanglement detection theoretically shown possible in \cite{Dasen15}.

Levitons can also be useful for new quantum detection schemes. For example, the Full Counting Statistics (FCS) of levitons partitioned by a quantum conductor has been theoretically shown detectable by sending synchronized levitons in a coupled Mach-Zehnder interferometer \cite{Dasen16}. Here, remarkably, no noise measurements are needed, only dc current measurements are necessary to access the electron FCS.

\begin{figure}[t]%
\includegraphics*[clip, width=7cm,height=6cm]{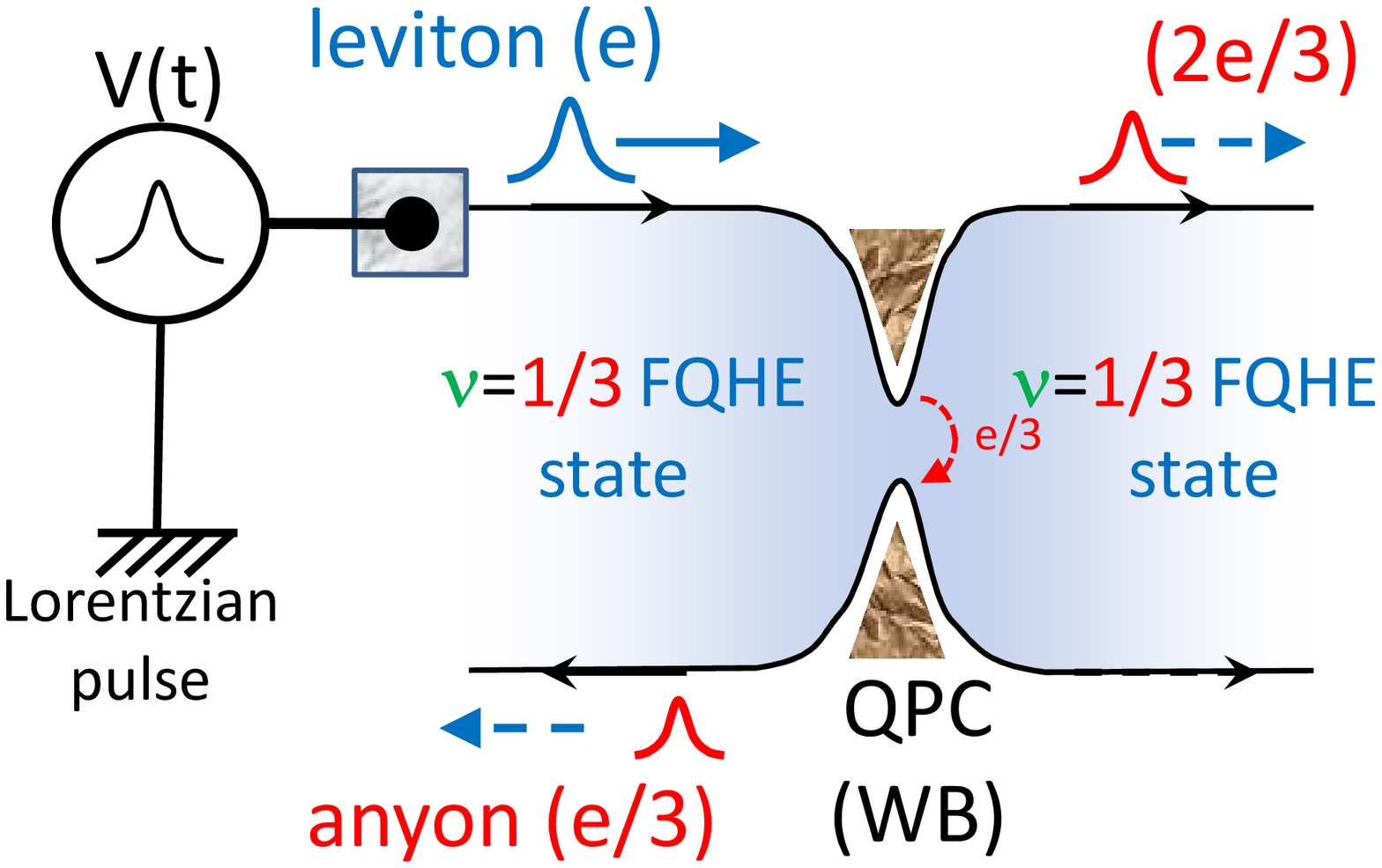}
\caption{%
  Proposal for a time-resolved anyonic source in the Fractional Quantum Hall regime (filling factor 1/3). Integer charge levitons are first created with a charge $e$ Lorentzian voltage pulse applied on the contact of a 1/3 fractional edge channel. Then, the leviton arrives at a Quantum Point Contact tuned in the weak backscattering (WB)  regime (weak electronic reflection probability). It is conjectured that the leviton will break into a $e/3$ backscattered and $2e/3$ transmitted fractionally charged pulses. From the backscattered side, this realizes a non-deterministic source of $e/3$ excitations with Poissonian statistics. The time resolved properties inherited from the original integer charge leviton may open the way for new quantum experiments to probe the anyonic statistics.}
\label{anyon}
\end{figure}

Regarding interaction,
it has been predicted that levitons can also be generated in an interacting 1D electron system \cite{Keel06} called a Luttinger Liquid (LL). A nice example is given by the chiral edge states of the Fractional quantum Hall effect (FQHE), whose best understood state, the Laughlin state, occurs at filling factor $1/3$ \cite{Laug83}. Here the current is carried by a single edge channel. This is a chiral Luttinger liquid which is expected to generate elementary excitations carrying fractional charge $e*=e/3$ \cite{Wen90} and showing anyonic statistics \cite{Arov84} in full correspondence with the bulk Laughlin excitations. This regime has been considered in \cite{Safi10} and more recently in \cite{Rech16}. Counterintuitively and contrary to what was suggested in \cite{Keel06}, no fractionally charged levitons can be generated in a fractional edge but only integer charge leviton. Indeed, it is mandatory that the total phase variation resulting from the voltage pulse be not a fraction of $2\pi$ in order to have only electron like excitations with no holes and thus a minimal excitation property. This implies that the Faraday flux, which in the 1/3 FQHE regime writes $\int e*V(t)dt$ be equal to $nh$. As $I(t)=e*(e/h)V(t)$ the total injected charge for $2n\pi$ phase shift is thus $q=ne$, where $n$ an integer. Can these integer charge minimal excitation states be used to probe the fascinating fractionally charge excitations? Having the time domain control of $e*$ excitations could open new way to probe their anyonic statistics via Hong Ou Mandel correlations or other two-particle interferometry protocol. In Fig.~\ref{anyon} we propose a possible source of on-demand anyons: integer charge levitons are first generated and then sent to a QPC which is tuned in the weak backscattering regime (i.e. shows a weak reflection probability). In this regime, shot noise measurements, using dc voltage source to create an incoming flux of integer charges, have shown that a Poissonian emission of fractional charges $e*$ can be emitted to contribute to the backscattering current \cite{Sami97},\cite{DePi97}. We conjecture that, similarly, an integer charge leviton would be partitioned into an $e/3$ backscattered charge and a $2e/3$ transmitted charge. We think this situation would lead to the generation of levitonic like charge pulses carrying fractional charge. Here, the deterministic emission of fractionally charge levitons would be lost because of the Poissonian backscattering statistics but the time resolution and the time domain control inherited from the initial integer charge levitons will be kept and could be used for further experiments like Hong Ou Mandel collision of anyons to probe their quantum statistics.

In another perspective it is interesting to address the physics of non-integer charge pulses. Only voltage pulses can realize this situation while quantum dot based charge sources can only emit integer charge particles. Indeed by continuously tuning the voltage pulse amplitude on a contact one can generate any value of the injected charge. It was shown from Eq.~\ref{fracpole} that only an integer $q$ gives rise to a minimal excitation state. Injecting fractional charges, even using Lorentzian pulses can not be viewed as pure levitons. However non-integer charge pulses created by Lorentzian voltage pulses have interesting properties which deserve attention. In \cite{Belz16} it was shown that half-levitons, i.e. charge e/2 created by Lorentzian voltage pulses, minimize the shot noise of a Superconducting/Normal conductor junction  the same way as integer charge levitons minimize the noise in a purely normal conductor. Here this occurs in the energy sub-gap regime where, while the normal current is suppressed by the superconducting gap, a small current is allowed via the so-called Andre'ev reflection mechanism. In Ref.~\cite{Mosk16}, the nature of half integer Lorentzian charge pulses have been studied and it has been shown that they can form remarkable fractionally charged zero-energy single-particle excitations states. Separating the half-leviton into a $e/2$ charge part and its accompanying neutral electron-hole cloud, \cite{Mosk16} showed that a half-leviton and a anti-half leviton mixed in a semi-transparent electronic beam-splitter can elastically annihilate, a property not shared with ordinary distinguishable electron and hole excitations.

Non-integer charge pulses behave differently than integer pulses in electronic Mach-Zehnder (MZI) or Fabry-Pérot (FPI) interferometers. This was first noticed in the work of \cite{Gaur14} where, from numerical dynamical simulations, it was remarked that, for pulses having an extension smaller than the MZI arm difference or the FPI perimeter, the transmission was dependent on the charge. For example, the Fabry-Pérot transmission shows oscillations with the charge $q$ modulo $e$. The explanation was found in the phase difference $2\pi (q/e)$, see Eq.~\ref{phaseincr}, between the front and the back of the charge wave-packet which combines with the orbital phase accumulated in the interferometer. Striking effect on the interference visibility of the current in an electronic Mach-Zehnder interferometer  have been also found in \cite{Hofe14}. For example, while the visibility as a function of an external magnetic flux vanishes when the injection period corresponds to have exactly one electron in the MZI for charge $q=e$, Ref.~\cite{Hofe14} found that for  $q=e/3$ the visibility vanishes when there are exactly three fractional charge pulses in the MZI. More generally robust visibility cancelation occurs when there are $p$ charges for fractional charges $q=ek/p$. Similar results are found in a FPI interferometer. This intriguing properties call for looking at possible anyonic or parafermionic quantum statistics that may obey these charge pulses.

The above mentioned perspectives are only a few direction and do not pretend to give a fair exhaustive account of the impressive literature that levitons have triggered and we apologize for missing references and topics. We think that we are only walking on the surface of a Fermi Sea Iceberg.

\begin{acknowledgement}
Support from the ERC Advanced Grant 228273 MeQuaNo is acknowledged.
\end{acknowledgement}

% Use the following code if you wish to generate your bibliography with BibTeX;
% replace the string "pss-demo" below with the name(s) of
% the BibTeX data base(s) you want to use.
% The resulting bibliography-output (the content of the .bbl file)
% must be pasted back into this file before submission.
% Please also include your BibTeX data base file(s) in your submission
% so that we can re-run BibTeX if necessary.
%
%\bibliographystyle{pss}
%\bibliography{pss-demo}
%
% Replace the following example bibliography with your references
% before submission:

\end{document}